\documentclass[aps,prd,superscriptaddress,twocolumn,preprintnumbers,nofootinbib,showpacs]{revtex4}

\usepackage{graphicx}
\usepackage{amsmath}

\newcommand{\MET}{{\slash\!\!\!\!E_T}}
\newcommand{\DZero}{D$\slash\!\!\!0$}

\begin{document}

\title{Top quark forward-backward asymmetry and $W^\prime$ bosons}

\author{Daniel~Duffty}
\affiliation{Illinois Institute of Technology, Chicago, Illinois 60616-3793, U.S.A.}

\author{Zack~Sullivan}
\email{Zack.Sullivan@IIT.edu}
\affiliation{Illinois Institute of Technology, Chicago, Illinois 60616-3793, U.S.A.}

\author{Hao Zhang}
\email{haozhang@anl.gov}
\affiliation{Illinois Institute of Technology, Chicago, Illinois 60616-3793, U.S.A.}
\affiliation{High Energy Physics Division, Argonne National Laboratory, 
Argonne, Illinois 60439, U.S.A.}

\begin{abstract}
  The top quark forward-backward asymmetry measured at the Fermilab
  Tevatron collider deviates from the standard model prediction. A
  $W^{\prime}$ boson model is described, where the coupling
  $W^{\prime}$-$t$-$d$ is fixed by the $t\bar t$ forward-backward
  asymmetry and total cross section at the Tevatron.  We show that
  such a $W^\prime$ boson would be produced in association with a top
  quark at the CERN Large Hadron Collider (LHC), thus inducing
  additional $t\bar t+j$ events.  We use measurements of
  $t\bar{t}+n$-jet production from the LHC to constrain the allowed
  $W^{\prime}$-$t$-$d$ couplings as a function of $W^\prime$ boson
  mass.  We find that this $W^{\prime}$ model is constrained at the
  $95\%$ C.L. using $0.7$~fb$^{-1}$ of data from the LHC, and could be
  fully excluded with 5~fb$^{-1}$ of data.
\end{abstract}

\preprint{IIT-CAPP-12-02, ANL-HEP-PR-12-15}


\maketitle

\section{Introduction}
\label{sec:intro}

Measurements of the forward-backward asymmetry in $t\bar t$ production
($A_{FB}^{t}$) by the CDF \cite{key-1} and \DZero\ \cite{key-2}
Collaborations have garnered great interest, as the experimental
results appear to disagree with standard model predictions
\cite{key-3} at the 95\% confidence level.  Many models have been
proposed to explain this anomaly as physics from beyond the standard
model (BSM).  Some models envision new $s$-channel
processes~\cite{key-s-channel_models} like axigluons, or other exotic
scenarios~\cite{key-other_models}.  Other models invoke new particles,
such as $W^\prime$ bosons~\cite{{key-wprime1,key-idea,key-wprime2,
    key-B-constraint,key-6,key-idea3}} or $Z^\prime$
bosons~\cite{key-zprime,key-zprime2,key-zprime3} that enter via
virtual $t$-channel exchange.  There are also model-independent ideas
about the $t\bar t$
asymmetry~\cite{key-model_independent,key-idea2,key-model_independent2}.

A new vector boson with a large flavor changing coupling between the
first and the third generation could induce a large enough $t\bar{t}$
charge asymmetry to explain the anomaly through a $t$-channel
exchange.  Neutral vector bosons ($Z^{\prime}$) are already
constrained by early Large Hadron Collider (LHC) data, as they would
produce too many same-sign top quark pair events at the LHC
\cite{key-zprime2,key-4}.  Hence, we focus in this paper on whether
charged vector currents ($W^\prime$ bosons) are a viable explanation
once faced with measurements from the LHC.

Standard model-like $W^\prime$ bosons are highly constrained by direct
measurements into final states with leptons at LHC ($m_{W^\prime}>
2.15$~TeV \cite{Aad:2011yg}) or with top quarks at the Fermilab
Tevatron ($m_{W^\prime}> 890$~GeV
\cite{Abazov:2011xs,Aaltonen:2009qu}).  To avoid any chance of direct
lepton production bounds, or flavor physics constraints
\cite{Nakamura:2010zzi}, we focus on right-handed $W^\prime$ bosons.
To avoid direct constraints from single-top-quark production, we
consider a non-standard $W^\prime$ which couples to one first and one
third generation right-handed quark.  The relevant interaction
Lagrangian for this model \cite{{key-wprime1,key-idea,key-wprime2,
    key-B-constraint,key-6,key-idea3}} may be written as
\begin{equation}
  \mathcal{L}=\frac{g}{\sqrt{2}}V^\prime_{td}\bar{d}\gamma^{\mu}P_{R}tW'_{\mu}
  +H.c.\;,\label{eqn:coupl}
\end{equation}
where $g$ is the standard model SU(2)$_{L}$ gauge interaction coupling
constant.  One could allow these $W^{\prime}$ bosons to couple strange
quarks to top quarks.  However, the strange quark parton density is
not large enough to meaningfully contribute to the $t\bar{t}$ charge
asymmetry.  Hence, we follow Refs.\
\cite{{key-wprime1,key-idea,key-wprime2,
    key-B-constraint,key-6,key-idea3}} and consider only the coupling
in Eq.~\ref{eqn:coupl} above.

While we can avoid direct production limits for $W^\prime$ bosons, in
this paper we demonstrate that early $t\bar t+j$ data from the
CERN Large Hadron Collider (LHC) already severely constrains these
models.  Hence, $W^\prime$ bosons are not expected to be a viable
solution to the $t\bar t$ forward-backward asymmetry anomaly.  We
organize the rest of this paper as follows: In Sec.\ \ref{sec:model}
we describe the analytic and numerical contribution of $W^\prime$
bosons to the $t\bar t$ forward-backward asymmetry, and derive the
parameters consistent with the Tevatron anomaly.  In Sec.\
\ref{sec:constraints} we examine the contribution of $W^\prime$ bosons
to $t\bar t+j$ measurements, set a 95\%~C.L. limit on the allowed
$W^\prime$ mass and coupling $V^\prime_{td}$, and exclude the
parameters required to explain the Tevatron anomaly.  We summarize our
results in Sec.\ \ref{sec:conclude}.

\section{$W^\prime$ model and $t\bar t$ asymmetry}
\label{sec:model}

The $W^\prime$ contribution to the $t\bar{t}$ production process
$d_{p_{1},\lambda_{1},c_{1}}+\bar{d}_{p_{2},\lambda_{2},c_{2}}\to
t_{p_{3},\lambda_{3},c_{3}}+\bar{t}_{p_{4},\lambda_{4},c_{4}}$ (where $p_i$,
$\lambda_i$, and $c_i$ are the four momentum, helicity, and color
factor of the $i$th particle, respectively), can be factorized into 
non-zero partonic level helicity amplitudes as
\begin{equation}
\displaystyle{\frac{g^{2}V_{td}^{\prime2}\hat{s}\delta_{c_{1}c_{3}}\delta_{c_{2}c_{4}}}{8(\hat{t}-m_{W'}^{2})}}\mathcal{M}(\lambda_{1},\lambda_{2},\lambda_{3},\lambda_{4}) \;.
\end{equation}
We have~\cite{key-12}
\begin{eqnarray}
\mathcal{M}(+,-,+,+) & = & -\sqrt{1-\beta^{2}}\left(2+r_{W^\prime}^2\right)\sin\theta,\\
\mathcal{M}(+,-,+,-) & = & -\left[2(1+\beta)+(1-\beta)r_{W^\prime}^2\right]\nonumber\\&&\times\left(1+\cos\theta\right),\\
\mathcal{M}(+,-,-,+) & = & \left[2(1-\beta)+(1+\beta)r_{W^\prime}^2\right]\nonumber\\&&\times\left(1-\cos\theta\right),\\
\mathcal{M}(+,-,-,-) & = & \sqrt{1-\beta^{2}}\left(2+r_{W^\prime}^2\right)\sin\theta,
\end{eqnarray}
where $\hat{s}=(p_{1}+p_{2})^{2}$, 
$\hat{t}=(p_{1}-p_{3})^{2}=-{\displaystyle \frac{1}{2}\hat{s}}\left(1-\beta\cos\theta\right)$,
$\beta=\sqrt{1-4m_{t}^{2}/\hat{s}}$, $r_{W^\prime}=m_{t}/m_{W^\prime}$ and 
$\theta$ is the angle between
$p_{1}$ and $p_{3}$ in the $t\bar t$ center of momentum frame.
The corresponding standard model (SM) helicity amplitudes are~\cite{key-12}
\begin{eqnarray}
\mathcal{M}(+,-,+,+) & = & -g_{s}^{2}t_{c_{2}c_{1}}^{a}t_{c_{3}c_{4}}^{a}\sqrt{1-\beta^{2}}\sin\theta,\\
\mathcal{M}(+,-,+,-) & = & -g_{s}^{2}t_{c_{2}c_{1}}^{a}t_{c_{3}c_{4}}^{a}\left(1+\cos\theta\right),\\
\mathcal{M}(+,-,-,+) & = & g_{s}^{2}t_{c_{2}c_{1}}^{a}t_{c_{3}c_{4}}^{a}\left(1-\cos\theta\right),\\
\mathcal{M}(+,-,-,-) & = & g_{s}^{2}t_{c_{2}c_{1}}^{a}t_{c_{3}c_{4}}^{a}\sqrt{1-\beta^{2}}\sin\theta.
\end{eqnarray}

The interference term $\sigma_{INT}$ between the SM amplitude
and the new physics amplitude is negative.  This property is useful in
explaining the Tevatron anomaly, because the interference term will
largely cancel the contribution to the $t\bar{t}$ inclusive cross
section from the new physics term.  As such, a $V^\prime_{td}$ can be
found that gives a large additional contribution to the
forward-backward asymmetry $A_{FB}^{t}$.

The total cross section of $t\bar{t}$ production at Tevatron is
$7.5\pm 0.48$~pb \cite{CDF:9913}.  The leading order (LO)
cross section obtained from MadEvent~5~\cite{key-madgraph5} is
$5.63$~pb using CTEQ6L1 parton distribution functions
(PDFs)~\cite{key-cteq}. The next-to-leading order (NLO) $t\bar{t}$
total cross section from MCFM~6.0~\cite{key-mcfm} is $6.9$~pb using
CTEQ6.6M PDFs (where the renormalization scale and factorization
scales are chosen to be $\mu_{r}=\mu_{f}=172.5$~GeV)~\cite{key-cteq}.
In order to address how new physics modifies the standard model
results, we rewrite the total SM cross section as
\begin{equation}
\sigma_{SM}^{NLO}=\sigma_{SM}^{NLO(F)}+\sigma_{SM}^{NLO(B)}=\sigma_{SM}^{LO}+\Delta\sigma_{SM}.
\end{equation}

The addition of $W^\prime$ bosons will modify the $t\bar t$ cross
section already at leading order, and lead to a ``new physics'' cross section
$\sigma_{NP}=\sigma_{SM}^{LO}+\sigma_{INT}+\sigma_{NEW}$.  We include
the NLO QCD correction in the SM part by considering
$\sigma_{NP}^{Tot}\equiv \sigma_{NP}-\sigma_{SM}^{LO}+\sigma_{SM}^{NLO}$.
The quantity of interest, $A_{FB}^{t}$, is calculated using
\begin{eqnarray}
A_{FB}^{t} & = & \frac{\sigma^{(F)}-\sigma^{(B)}}{\sigma^{(F)}+\sigma^{(B)}}\nonumber\\
 & = & \frac{\sigma_{NP}^{(F)}-\sigma_{SM}^{LO(F)}+\sigma_{SM}^{NLO(F)}}{\sigma_{NP}-\sigma_{SM}^{LO}+\sigma_{SM}^{NLO}}\nonumber\\
 &&-\frac{\sigma_{NP}^{(B)}-\sigma_{SM}^{LO(B)}+\sigma_{SM}^{NLO(B)}}{\sigma_{NP}-\sigma_{SM}^{LO}+\sigma_{SM}^{NLO}}\nonumber\\
 & = & \frac{\sigma_{NP}^{(F)}+\sigma_{SM}^{NLO(F)}-(\sigma_{NP}^{(B)}+\sigma_{SM}^{NLO(B)})}{\sigma_{NP}-\sigma_{SM}^{LO}+\sigma_{SM}^{NLO}}\nonumber\\
 & = & \frac{(\sigma_{NP}^{(F)}-\sigma_{NP}^{(B)})+(\sigma_{SM}^{NLO(F)}-\sigma_{SM}^{NLO(B)})}{\sigma_{NP}-\sigma_{SM}^{LO}+\sigma_{SM}^{NLO}}\nonumber\\
 & = & \frac{A_{FB}^{NP}\times\sigma_{NP}+A_{FB}^{SM}\times\sigma_{SM}^{NLO}}{\sigma_{NP}^{Tot}},
\end{eqnarray}
where $A_{FB}^{SM}=5.0\%$ \cite{key-2}, and
$\sigma_{NP}^{(F)}-\sigma_{NP}^{(B)}$ is obtained from events
generated using MadEvent~5.  The NLO QCD correction to the
$W^{\prime}$ model can be found in Ref.~\cite{key-6}, however, that
work found it to be numerically small (at most a few percent), and so
we do not include it in this work.

\begin{figure}
\includegraphics[scale=0.3]{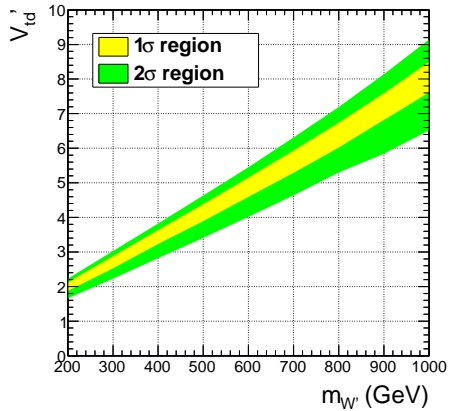}
\caption{Region of $W^\prime$ coupling $V_{td}^\prime$ vs.\ $W^\prime$
  mass consistent with Tevatron measurements of the $t\bar t$
  asymmetry.\label{fig:The-allowed-region}}
\end{figure}

In order to establish the relevant parameters of the $W^\prime$ model,
we scan the region [200 GeV, 1000 GeV]$\times$[0.1, 10.0] in the
parameter space ($m_{W^{\prime}}$, $V^\prime_{td}$) using 
a generic $W^\prime$ model file \cite{ZhouSullivan} in 
MadEvent~5
with CTEQ6L1 PDFs and a floating scale scheme.  The $W^{\prime}$ width
is given by 
\cite{key-5}
\begin{equation}
\Gamma_{W'}=\frac{g^{2}|V_{td}^{\prime}|^2m_{W'}}{16\pi}\left(1-r_{W^\prime}^2\right)\left(1+\frac{r_{W^\prime}^2}{2}\right) \;.
\end{equation}
The width is narrow, and is checked using BRIDGE~\cite{key-bridge}.
The $t\bar{t}$ asymmetry is compared with the unfolded result from
\DZero\ \cite{key-2}.  The unfolded result is obtained with the
assumption that the modified event distribution is the same as in the
standard model. This is not exactly correct for our new physics model,
but the difference is found to be small.\footnote{We confirm the
  claims of Ref.\ \protect\cite{key-6} that the change in acceptance
  between the standard model and $W^\prime$ model is small.}
Neglecting the non-trivial correlation between $\sigma_{NP}^{Tot}$ and
$A_{FB}^{t}$, for simplicity, we do a combined fit to both variables.
The $1\sigma$ ($2\sigma$) regions of allowed parameter space are
defined by
\begin{equation}
\frac{\left(\sigma_{t\bar t}-\sigma_{t\bar t}^{\rm obs}\right)^2}
{\delta\sigma_{t\bar t}^2}+\frac{\left(A_{FB}-A_{FB}^{\rm obs}\right)^2}
{\delta A_{FB}^2}\leq 1(4)\;,
\label{sigma}
\end{equation} 
and shown in Fig.~\ref{fig:The-allowed-region}.  These regions are
consistent with the full NLO results of Ref.\ \cite{key-6}.

\section{Constraints from the LHC}
\label{sec:constraints}

\begin{figure}
\includegraphics[scale=0.5]{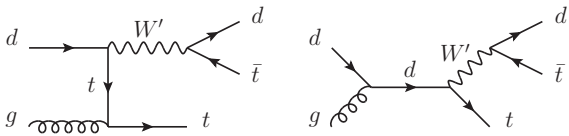}

\caption{The Feynman diagrams of $t\bar{t}+j$ production in this
  $W^\prime$ model.\label{fig:feyn_diag}}
\end{figure}

While $s$-channel production of a $W^\prime$ boson is explicitly
turned off in this model, the $W^{\prime}$ boson could be produced in
association with a top quark at the LHC.  The final state will be
$t\bar{t}+j$ (see Fig.\ \ref{fig:feyn_diag}).  This signal can easily
be checked at the LHC~\cite{key-idea,key-idea2,key-idea3}.  Both the
ATLAS \cite{key-7,key-8,key-9} and CMS \cite{key-10} collaborations
have published results of the inclusive and $t\bar{t}+n$-jet cross
section measurements.

The strongest constraint on our model comes from the dilepton decay
mode of top quark pair production measured by ATLAS \cite{key-7}
using an integrated luminosity of $0.70$~fb$^{-1}$.  The topology of
the final state is an opposite-sign dilepton pair with three jets and
large missing transverse energy $\MET$.  We simulate detector effects
by smearing jets and leptons with an energy resolution parametrized by
${\displaystyle \frac{\delta E}{E}=\frac{a}{\sqrt{E}}\oplus b}$; where
$a=0.5$, $b=0.03$ for jets~\cite{key-sextet}, $a=0.1$, $b=0.02$ for
electrons~\cite{key-sextet,key-elec}, and $a=0.04$, $b=0$ for
muons~\cite{key-muon}.  We calculate the missing transverse energy
$\MET$ after smearing from the imbalance of the reconstructed jets and
leptons.  To compare with the ATLAS $t\bar t+j$ analysis, we add cuts
on the smeared events as follows:
\begin{itemize}
\item Electrons: $p_{T e}>25$~GeV, $|\eta_e|<1.37$
  or $1.52<|\eta_e|<2.47$;
\item Muons: $p_{T \mu}>20$~GeV, $|\eta_\mu|<2.5$;
\item Jets: $p_{T j}>25$~GeV, $|\eta_j|<2.5$;
\item $\Delta R_{jj}>0.4$, $\Delta R_{ej}>0.4$, $\Delta R_{\mu j}>0.4$,
$\Delta R_{\mu\mu}>0.3$, $\Delta R_{e\mu}>0.2$, $\Delta R_{ee}>0.2$;
\item and the invariant mass of the charged leptons $m_{ll}>15$~GeV.
\end{itemize}
After acceptance cuts, different cuts are added to $ee$ and $\mu\mu$,
or $e\mu$ events.
\begin{itemize}
\item For $ee$ and $\mu\mu$ events, the missing transverse energy
  $\MET>60$~GeV, and $m_{ll}$ must differ by at least
  10~GeV from the $Z^0$-boson mass.
\item For $e\mu$ events, the scalar sum of the transverse momenta of
jets and leptons $H_{T}>130$~GeV.
\end{itemize}

We compare our result with the ATLAS data shown in Figure 1(a) of
Ref.\ \cite{key-7}.  There will be a contribution from higher order
corrections to $tW^{\prime}+$jets if some of the partonic jets are
merged by the jet reconstruction algorithm.  The $tW^{\prime}$ process
could also be detected in events with more than three jets due to
initial state radiation (ISR) and final state radiation (FSR).  To
mimic these effects on acceptance, we rescale our calculation by
comparing our SM $t\bar t+j$ results from MadEvent~5 (with cuts and
smearing) to the theoretical prediction (after cuts) used in Ref.\
\cite{key-7}.  All of the new physics results are rescaled by this
same factor and then compared with the data.  We note that the
observed event number by ATLAS is a little larger than the SM
prediction, which slightly weakens the constraint we extract from the
data.

\begin{figure}
\includegraphics[scale=0.3]{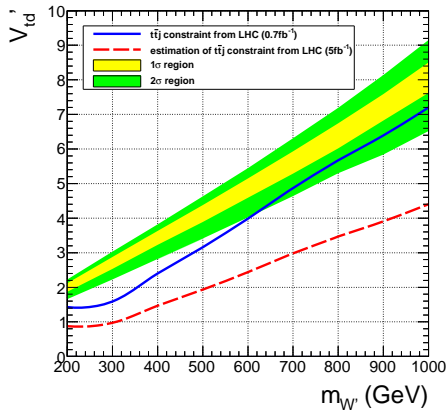}

\caption{Constraint from the LHC $t\bar{t}+j$ search on
  the $W^{\prime}$boson.  The parameter space above the solid blue line is
  excluded by the ATLAS data with $0.7\mathrm{\ fb}^{-1}$ of integrated
  luminosity at a $2\sigma$ level.  We also show the expected exclusion
  curve (the dashed red line) with $5\mathrm{\ fb}^{-1}$ of integrated
  luminosity.\label{fig:The-constraint}}
\end{figure}

In Fig.~\ref{fig:The-constraint} we show the allowed parameter space
consistent with the Tevatron forward-backward asymmetry anomaly, and
the independent $2\sigma$ bound on $V_{td}^\prime$ we extract from the
fit to ATLAS data.  We see that already with the first $0.7$~fb$^{-1}$
data, the $1\sigma$ region of parameter space consistent with the
Tevatron $A_{FB}^t$ is completely excluded at greater than a $95\%$
confidence level (C.L.).  Below 600 GeV the $2\sigma$ region of
parameter space is also excluded at 95\% C.L..

In the process we are examining, $\sigma\left(pp\to tW^{\prime}\to
  t\bar{t}d\right)\propto V_{td}^{\prime2}$, the cross section
significance $S/\sqrt{B}$ scales like $\sqrt{N}$, where $N$ is event
number.  Hence, the bound on $V_{td}^\prime$ will decrease
$\propto\mathcal{L}^{-1/4}$ when the integrated luminosity
$\mathcal{L}$ increases.  We use this scaling to estimate the bound on
$V_{td}^\prime$ that can be reached with the existing 5~fb$^{-1}$ of 
integrated luminosity, and show this bound in
Fig.~\ref{fig:The-constraint} (the dashed red line).  With 5~fb$^{-1}$
data, a $W^{\prime}$ model with coupling constant $V_{td}^\prime$
large enough to explain the Tevatron top quark forward-backward
asymmetry anomaly can be unambiguously excluded.

\begin{figure}
\includegraphics[scale=0.3]{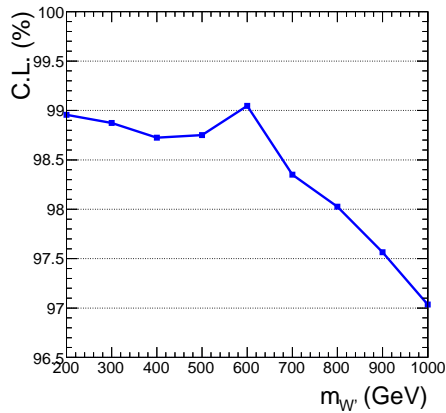}
\caption{Exclusion level (in percent) of the $W^{\prime}$ model
from a simultaneous fit of three experimental observables.
  \label{fig:The-exclusion-level}}
\end{figure}

In addition to considering the independent limit on $W^\prime$
production from LHC data, we also consider the limit obtained by a
combined fit to the $t\bar t$ total cross section at the Tevatron,
$A_{FB}^{t}$, and $t\bar{t}+j$ from the LHC.  Since there are 2 free
parameters ($m_{W^\prime},V_{td}^\prime$), we have
\begin{eqnarray}
  \chi^2/\mathrm{d.o.f.}&=&\frac{1}{3-2}\biggl[\frac{\left(\sigma_{t\bar t}
      -\sigma_{t\bar t}^{\mathrm{Tev}}\right)^2}
  {\delta\sigma_{t\bar t}^2}+\frac{\left(A_{FB}-A_{FB}^{\mathrm{Tev}}\right)^2}
  {\delta A_{FB}^2}\nonumber\\
  &&+\frac{\left(\sigma_{t\bar tj}-\sigma_{t\bar tj}^{\mathrm{LHC}}\right)^2}
  {\delta\sigma_{t\bar tj}^2}\biggr]
\end{eqnarray}
for the right-handed $W^\prime$ model.  The confidence region is
calculated from the $\chi^2$ cumulative distribution with 1 degree of
freedom.  The result is shown in Fig.~\ref{fig:The-exclusion-level}.
A simultaneous fit excludes a right-handed $W^\prime$ model at more
than a $97\%$ confidence level (C.L.).  While $A_{FB}^t$ provides
tension with the standard model at the Tevatron, a simultaneous fit
for all three measurements is only excluded at the $92\%$ C.L..  In
other words, the standard model agrees better with data than the
attempted $W^\prime$ boson fix.

\section{Conclusions}
\label{sec:conclude}

We study a right-handed $W^{\prime}$ model which has been suggested as
an explanation of the Tevatron $t\bar t$ forward-backward asymmetry
anomaly in the context of recent measurements from the Large Hadron
Collider.  Measurements of inclusive $t\bar{t}$ production constrain
this $W^{\prime}$ model, because a $W^{\prime}$ boson would induce
extra $t\bar{t}+j$ events.  We find that the values of the $W^\prime$
mass and coupling $V_{td}^\prime$ required to fit both
$\sigma_{t\bar{t}}$ and $A_{FB}^{t}$ at Tevatron at the $2\sigma$
level, are excluded at $95\%$ C.L.\ by measurements of $t\bar t j$
with $0.7$~fb$^{-1}$ of data by the ATLAS Collaboration.  If the full
5~fb$^{-1}$ data set is analyzed, the measurement of $t\bar tj$ alone
will push this limit to more than $3\sigma$.  We also show that a
simultaneous fit to three measurements excludes $W^\prime$ bosons as an
explanation for the Tevatron $t\bar t$ forward-backward asymmetry
at a $97\%$ C.L..

In addition to the measurements considered here, we point out the
\DZero\ Collaboration measures the charge asymmetry of the
charged leptons ($A_{FB}^l$) from top quark decay in $t\bar t$
events~\cite{key-2}.  Due to angular correlations between the top
quark and the charged lepton from its decay, it has been shown that
there is a correlation between $A_{FB}^t$ and
$A_{FB}^l$~\cite{key-12} that suggests a light right-handed
$W^\prime$ boson is preferred by the data.  The limits we obtain from
the LHC with $0.7$~fb$^{-1}$ of data are even stronger for light
$W^\prime$ bosons (nearly 99\% C.L. exclusion) than for heavier
$W^\prime$ bosons.  Adding $A_{FB}^l$ information from \DZero\
would further disfavor this $W^\prime$ boson model.

We conclude by noticing that even though the $W^\prime$ boson only
couples to the right-handed top and down quarks, there are still
constraints from flavor physics.  The constraint from $B\to \pi K$ is
strong, and the right-handed $W^\prime$ model here may also be
constrained by the branching ratio of rare $B$ decays at the $2\sigma$
level~\cite{key-B-constraint}.  However, due to a relatively large
theoretical uncertainty for the $B$ decays (even for the standard
model prediction~\cite{key-rare-B}), the direct production limit we
present from collider physics is needed to exclude this right-handed
$W^\prime$ model.

\begin{acknowledgments}
  This work is supported by the U.~S.\ Department of Energy under
  Contract No.\ DE-FG02-94ER40840. Additional support for H.~Zhang is
  provided under Contract No.\ DE-AC02-06CH11357.
\end{acknowledgments}

\end{document}